\documentclass[a4paper,twoside,12pt]{article}
\usepackage{xypic}
\usepackage{a4}
\usepackage{amsmath}
\usepackage{amssymb} 

\DeclareMathSymbol{\ltimes} {\mathbin}{AMSb}{"6E}
\DeclareMathSymbol{\rtimes} {\mathbin}{AMSb}{"6F}

\newtheorem{theorem}{Theorem}[section] 
\newtheorem{lemma}[theorem]{Lemma} 
 
\newtheorem{corollary}[theorem]{Corollary} 
\newtheorem{algorithm}[theorem]{Algorithm}
\newtheorem{definition}[theorem]{Definition}
\newtheorem{remark}[theorem]{Remark}

\renewcommand{\rhd}{\vartriangleright}
\renewcommand{\lhd}{\vartriangleleft}
\renewcommand{\Box}{\square}
\newcommand{\F}[0]{\mathbb{F}}
\newcommand{\DFT}[0]{{\rm DFT}}

\newcommand{\z}[0]{{\sf Z}}

\def\bra#1{\left<#1\right|}
\def\ket#1{\left|#1\right>}

\begin{document}

%
%

\begin{titlepage}

\begin{center}
{\Large Polynomial-Time Solution to the Hidden Subgroup Problem
for a Class of non-abelian Groups\par}

\vspace*{1cm}

\begin{center}
\begin{tabular}{c@{\qquad}c}
Martin R\"otteler\footnotemark[1] & Thomas Beth\\
{\tt roettele\symbol{64}ira.uka.de} & 
{\tt EISS\_Office\symbol{64}ira.uka.de}
\end{tabular}
\end{center}
\
\footnotetext[1]{supported by DFG grant GRK 209/3-98}

\vspace*{1cm}
Insitut f\"ur Algorithmen und Kognitive Systeme\\
Universit\"at Karlsruhe, Germany\\

\vfill
{\bf Abstract}

\bigskip
\begin{minipage}{12cm}
\noindent
We present a family of non-abelian groups for which the hidden
subgroup problem can be solved efficiently on a quantum computer.
\end{minipage}

\vfill
\end{center}
\end{titlepage}

%
%

\section{Introduction}
The hidden subgroup problem has found recent interest in the theory of
quantum computing. This is due to the fact that the power of quantum
computation compared to classical computation becomes apparent in this
problem.

The first occurrence of a hidden subgroup problem for an abelian group
appeared has been implicitly in Simon's work \cite{Simon:94}. In fact
he solved the hidden subgroup problem for the group $\z_2^n$ under the
promise that the hidden subgroup is of order $2$. The group
theoretical interpretation of this algorithm has been formulated by
several authors (see \cite{BH:97}, \cite{Jozsa:98}, \cite{ME:98}). In the
paper \cite{BH:97} it is shown that subgroups of arbitrary order can
be found and furthermore that this can be done by an exact quantum
polynomial time algorithm. Thus there is an exponential speed-up of
this quantum algorithm over any classical algorithm, even
probabilistic ones.

Recently the question has been raised as to whether the hidden
subgroup problem could also be solved for non-abelian groups. In the
paper \cite{EH:98} the problem is addressed for the dihedral groups
$D_N$. The authors have found an interesting way to circumvent the
application of the Fourier transform for the dihedral groups and
instead use the Fourier transform for $\z_N \times \z_2$ and still
learn something from the probability distribution about the existence
or non-existence of certain elements. However the classical
post-processing requires an optimization problem which makes the
overall algorithm exponential in the number of classical steps, but
is polynomial in the number of evaluations of the quantum black-box
circuit representing the given function.

In this paper we will present a family $W_n$ of non-abelian groups for
which the hidden subgroup problem can be solved by a number of steps
polynomial in the number of qubits.  The groups in this family are
certain semi-direct products (namely wreath products) and have some
desirable properties (they are, e.\,g., of bounded exponent).
Moreover, the fact that a Fourier transform for $W_n$ can be performed
efficiently by a quantum computer is important to solve the hidden
subgroup problem for these groups.
\bigskip

%
%

\section{The Hidden Subgroup Problem}

We adopt the definition of the hidden subgroup problem given in
\cite{EH:98}. The history of the hidden subgroup problem parallels the
history of quantum computing since the algorithms of Simon
\cite{Simon:94} and Shor \cite{Shor:94} can be formulated in the
language of hidden subgroups (see e.\,g. \cite{Jozsa:98} for this
reduction) for certain abelian groups. In the paper \cite{BH:97} an
exact quantum algorithm (running in polynomial time in the number of
evaluations of the given black box function and the classical
post-processing) is given for the hidden subgroup problem in the
abelian case. 

\begin{definition}[The hidden subgroup problem]\ \\
 Let $G$ be a finite group and $f:G \rightarrow R$ a mapping from $G$
to an arbitrary domain $R$ fulfilling the following conditions:

\begin{itemize}
\item[a)] The function $f$ is given as a quantum circuit, i.\,e., $f$
can be evaluated in superpositions.
\item[b)] There exists a subgroup $U \subseteq G$ such that $f$ takes a
constant value on each of the cosets $g U$ for $g \in G$.
\item[c)] Furthermore $f$ takes different values on different cosets.
\end{itemize}

\noindent
The problem is to find generators for $U$.
\end{definition}

%
%

\section{Wreath Products}

In this section we recall the definition of wreath products in general
(see also \cite{HuppertI:83} and \cite{JK:82}) and define the family
of groups for which we will solve the hidden subgroup problem.

\begin{definition} Let $G$ be a group and $H \subseteq {\rm S}_n$ be a
subgroup of the symmetric group on $n$ letters.
The wreath product $G \wr H$ of $G$ with $H$ is the set
\[ 
\{ (\varphi, h) : h \in H, \varphi: [1,\dots,n] \rightarrow G\}
\]
equipped with the multiplication
\[ 
(\varphi_1, h_1) \cdot (\varphi_2, h_2) := (\psi, h_1 h_2),
\]
where $\psi$ is the mapping which sends $i \mapsto \varphi_1(i^{h_2})
\varphi_2(i)$ for $i \in [1,\dots,n]$.
\end{definition}

The wreath product is isomorphic to a semidirect product of the
so-called {\em base group} $N := G\times \ldots \times G$ which is the
$n$ fold direct product of (independent) copies of $G$ with $H$, in
symbols $G \wr H = N \rtimes H$, where $H$ operates via permutation of
the direct factors of $N$. So we can think of the elements to be
$n$--tuples of elements from $G$ together with a permutation $\tau$
and multiplication is done component-wise after a suitable permutation
of the first $n$ factors
\[ (g_1, \dots, g_n; \tau) \cdot (g^\prime_1, \dots, g^\prime_n;
\tau^\prime) = (g_{\tau^\prime(1)} g^\prime_1, \dots, 
g_{\tau^\prime(n)} g^\prime_n; \tau \tau^\prime).
\]

%
%

\section{The Wreath Products $\z_2^n \wr \z_2$}

In the following we show some elementary properties of $W_n := \z_2^n
\wr \z_2$. The groups $W_n$ have exponent $4$ and base-group $N :=
\z_2^n \times \z_2^n$. Elements of $W_n$ are denoted by $(x, y; a)$
where $x, y\in \z_2^n$ and $a \in \z_2$. For a subgroup $U$ and an
element $g\in W_n$ as usual we define $U^g := \{ g^{-1} u g : u \in U
\}$. We think of the elements encoded in such a way that $x$ and $y$
are encoded in the lower significant bits and $a$ is the most
significant bit.
We later need the important
\begin{lemma}
Let $U$ be a subgroup of $W_n$ and $t=(0,0;1)$. Then
\begin{equation}\label{canonicfactor}
  U = (U \cap N) \cdot (U \cap U^t).
\end{equation}
\end{lemma}
{\bf Proof:} "$\supseteq$" is clear since $U \cap N$ and $U\cap U^t$
are subgroups of $U$. \\
"$\subseteq$": Let $u\in U$ be a given element. Since $u\in N$ implies
that $u$ is contained in the left factor, we can assume that
$u\notin N$, i.\,e., $u=(x, y; 1)$ for certain $x, y \in \z_2^n$. 
We compute 
\[ u^2 = (x\oplus y, x\oplus y;0), \quad u^3=(y, x;1), \quad
u^4=(0,0;0).
\]
Thus the effect of conjugating $u$ with $t$ is $u^t = (y,x;1)$ from which
we can deduce $u\in U^t$ since $u=(u^3)^t\in U^t$.\hfill $\Box$

\begin{remark}
\rm 
\begin{itemize}
\item[a)]
The preceding lemma shows that each subgroup $U$ of $W_n$ factorizes
in a canonical way into the product of two subgroups. Therefore it is
sufficient to find generators for $U\cap N$ and $U \cap U^t$ to obtain
a set of generators for $U$.
\item[b)] The action on an element $n=(x^\prime,y^\prime; 0) \in N$ of
an arbitrary transversal element $\tau = (x, y;1)$ for which
\[ W_n \stackrel{\{1, \tau\}}{\rhd} N \rhd E
\]
holds, is given by 
\[\tau^{-1}
n \tau = (y, x;1) (x^\prime, y^\prime; 0) (x, y;1) = (y, x;1) 
(x\oplus y^\prime, y \oplus x^\prime; 1) = (y^\prime, x^\prime;0),
\]
i.\,e., the components of $n$ are swapped.
\item[c)] From the isomorphism theorem for $U\subseteq W_n$ follows
\[ NU/N \cong U/U \cap N.
\]
Thus the index of $U \cap N$ in $U$ can be $1$ or $2$, since $N$ is a
maximal normal subgroup of $W_n$. So the subgroup $U\cap U^t$ indicates
whether the index is $1$ or $2$ and the case of index $2$ occurs 
iff there exists an element of the form $(x, y;1)$ in $U$.
\item[d)] If the index $[U : U\cap N] = 2$ then $(U\cap N)^t = U\cap
N$. This is readily seen from b) by observing the fact that there must
exist an element of the form $(a, b; 1) \in U$.
\item[e)] We call subgroups fulfilling the property $U = U^t$ {\bf
balanced}. Later on the balanced subgroups of $U$ will play an
important r\^ole since they will appear naturally in the process of
sampling.
\end{itemize}
\end{remark}

\begin{figure}[bht]
\[\xymatrix{
& W_n \ar@{-}[dl] \ar@{-}[ddr] &  \\
N \ar@{-}[ddr] & & \\
&& D \ar@{-}[dl] \\
& N\cap D  \ar@{-}[d]& = \zeta(W_n) = (W_n)^\prime = \Phi(W_n)\\
& E &
}
\]
\caption{\label{factorization}The wreath product $W_n$ factors over
$N$ and $D$}
\end{figure}
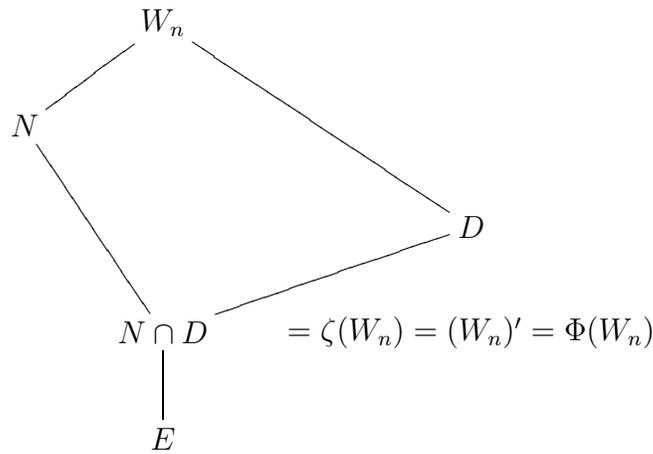

\subsection{Finding Involutions in $W_n$}
In this section we present a straightforward method to find hidden
subgroups of order $2$ in $W_n$; i.\,e., if it is promised that
$U$ has order $2$, then the generator of $U$ can be found without
invoking non-abelian Fourier transforms.

Consider the restriction of $f$ to the base group $\z_2^n \times
\z_2^n$: We can obtain a equal distributed superposition over the base
group $N$ by application of the $2\times 2$ Hadamard matrix $H$ on all
qubits except for the most significant one.

Then we use the quantum algorithm for the hidden subgroup problem for
$\z_2^{2 n}$ using only the first $2n$ bits. This gives generators for
the group $U \cap N$, so that the number of evaluations of $f$ is
linear in $n$ (also the classical post-processing needs only a number
of steps which is linear in $n$).

Note that each element in $N$ has order $2$ but there may be more
involutions in $W_n$. More precisely:
Each involution is contained in $N$ or $D$ where $D$ is defined by
\[ D = \{ (x, x; a), \; \mbox{where} \; x\in \z_2^n \; \mbox{and} \;
          a \in \z_2 
       \}
\]
This is because of the observation that for an element $(x, y; a) \in
W_n$
\[ 
(x, y; a)^2 = (x \oplus y, x \oplus y, 0) \stackrel{!}{=} (0,0;0)
\Rightarrow x = y
\]
holds. We also denote $D$ by $(\z_2^n \| \z_2^n) \times \z_2$
since the two factors are diagonal.

Figure \ref{factorization} shows the situation involving $W_n$, $N$
and $D$. Interestingly, the intersection of $N$ and $D$ is the center
$\zeta(W_n)$ of $W_n$ which coincides with the commutator
${W_n}^\prime$ and the Frattini subgroup $\Phi(W_n)$ of $W_n$ but
these facts will not be used in the sequel.

\begin{figure}[hbt]
\input{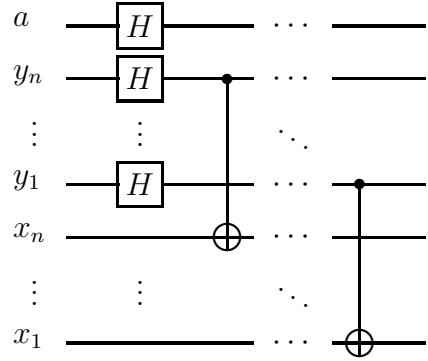}
\vspace*{0.3cm}
\caption{\label{involutions}Finding involutions in $D$}
\end{figure}

Since $D$ is abelian we can solve the hidden subgroup problem for $D$
by the usual abelian hidden subgroup algorithm. We do this by
performing Hadamard transforms on the qubits representing $y$ and $a$
followed by controlled NOTs between qubits $x_i$ and $y_i$ for
$i=1,\ldots, n$ (see figure (\ref{involutions})). This is followed by
the hidden subgroup algorithm for $\z_2^{n+1}$ applied to the bits
representing $y$ and $a$.

\subsection{The Pairing on $W_n$}\label{pairing}

In view of the Fourier transform for $W_n$ to come we define
(mimicking the abelian case) a pairing $\mu$ on $W_n$ which in turn
allows the definition of "´duals"´ needed to treat this case of
non-abelian groups.

\begin{definition}\label{defpairing}
We denote by $\mu:W_n \times W_n \rightarrow \z_2$ the pairing
\renewcommand{\arraystretch}{1.5}
\[ \mu((x, y;a), (x^\prime, y^\prime; a^\prime)) 
:= \left\{ \begin{array}{c@{\quad:\quad}l}
\sum x_i x^\prime_i + \sum y_i y^\prime_i & a = a^\prime = 0\\
\sum x_i y^\prime_i + \sum x^\prime_i y_i & 
a \oplus a^\prime = 1 \\
\sum x_i x^\prime_i + \sum y_i y^\prime_i + 1 & a = a^\prime = 1\\
\end{array}\right.
\]
\renewcommand{\arraystretch}{1}
\end{definition}

Like in the abelian case we denote suggestively the set of
perpendicular elements for a given $U\subseteq W_n$ by 
\[U^\perp := \{ g \in W_n : \forall h\in U: \mu(g, h) = 0. \}
\]

However, in general $U^\perp$ will not be a group any more. We will
also make use of a bijective mapping (which is of course not a homomorphism) 
$\varphi : W_n \rightarrow \F_2^{2n+1}$ which sends
\[ \varphi: \begin{array}{rcl} (x, y; 0) & \mapsto & (x, y, 0)\\
                               (x, y; 1) & \mapsto & (y, x, 1)
            \end{array}, \quad \mbox{where} \quad x, y\in \z_2^n. 
\]
Using $\varphi$ we can compute $\mu$ using the identity
\[ \forall g, h \in W_n : \mu(x, y) = 
\langle \varphi(g), \varphi(h) \rangle_{\F_2^{2n +1}}
\]
which holds due to the construction of $\mu$ and $\varphi$.

Here and in the following we let $t$ denote the element $(0,0;1)\in
W_n$. 
\begin{lemma}\label{perp}
For a subgroup $U$ of $W_n$ the following holds:
\[ y \not\perp U \Rightarrow \sum_{x \in U} \mu(x, y) = 0.
\]
\end{lemma}

\noindent
{\bf Proof:} Write $U=(U\cap N)\cdot (U\cap U^t)$. Two cases can
occur: \\
1.) $U=U\cap N$ (i.\,e. $U$ is abelian). Then
\[ \varphi(U) = \{ (x_i, y_i, 0) \in \F_2^{2n+1}: \;(x_i, y_i;0) \in U \}.
\]
This is a linear subspace of $\F_2^{2n+1}$ and since $y\not\perp U$
there exists $v\in \varphi(U)$ such that $\langle\varphi(y),
v\rangle\not= 0$. Invoking an $\F_2$-vector space argument
we conclude $\sum_{u \in \varphi(U)} \langle\varphi(y), u\rangle =
0$.\\ 2.) $U = (U\cap N) \stackrel{\cdot}{\cup} (U \cap N)\cdot t_0$
with $t_0 = (a, b;1)$. Due to the preceding remark we have $(U\cap
N)^t = (U\cap N)$, i.\,e., $\{(x_i, y_i;0)\in U\cap N\} = \{(y_i,
x_i;0)\}$ and mapping via $\varphi$ yields the decomposition of
$\varphi(U)$ into a vector space $V$ and an affine space
\[ \varphi(U) = \underbrace{\{(x_i, y_i;0)\}}_{=:V}
\stackrel{\cdot}{\cup}
(b, a, 1) \oplus \underbrace{\{(y_i, x_i;0)\}}_{=V},
\]
since $V$ is a balanced.

If there exists, an element $x_0 \in V$ with $y\not\perp x_0$ then
$\sum_{x\in V}\langle\varphi(y), x\rangle = 0$ and also $\sum_{x\in
V}(\langle\varphi(y), (b,a,1)\rangle + \langle\varphi(y), x\rangle) =
0$.

If no such element exists $y$ is perpendicular on $V$ and therefore
necessarily $\langle\varphi(y), (b, a, 1)\rangle = 1$. This means that
$\sum_{x\in V}(\langle\varphi(y), (b,a,1)\rangle + \langle\varphi(y),
x\rangle) = \sum_{x\in V}\langle\varphi(y), (b,a,1)\rangle = 0$.\hfill
$\Box$

We state another useful property of $\mu$ which will be needed later
on.

\begin{lemma}\label{halves}
Let $U$ be a subgroup of $W_n$. If there exists an element of the form
$(x, y; 1) \in U^\perp$ then exactly half of the elements of $U^\perp$
are in $N$.
\end{lemma}
{\bf Proof:} This follows from the fact that for $g\in W_n$, $h=(x,
y;1)$ and $u=(x^\prime, y^\prime; 0)$ we have:
\[ \mu(g, u \cdot h) = \mu(g, u) \oplus \mu(g, h),
\]
which follows from an easy computation. \hfill $\Box$

\subsection{The Lattice of Balanced Subgroups}
We have introduced the pairing $\mu$ on $W_n$ with respect to which we
can define orthogonal complements. However, as stated, for a given $U$ the
orthogonal complement $U^\perp$ need not again be a group. 

For example in case of $W_1$ we have 
\[ U = \{ (0,0;1), (0,1;0) \}, \quad 
U^\perp = \{ (0,0;0), (0,0;1), (0,1;1), (1,0;0)\}
\]
and $(0,1;1)^2 = (1, 1;0) \notin U^\perp$.
But we have the following
\begin{theorem}\label{balancedness}
Let $U\subseteq W_n$ be a subgroup and $t=(0,0;1)$. Then
\[ U = U^t \Leftrightarrow U^\perp \; \mbox{is a subgroup of} \; \; W_n.
\]
\end{theorem}

\noindent
{\bf Proof:} "$\Rightarrow$": By looking at the linear equations
defining $U^\perp$ when we employ the bijection $\varphi$, we firstly
observe that $U^\perp$ is again balanced:
\begin{equation}\label{keyBalanced}
\left[\begin{array}{c} \vdots \\ x_i, y_i, a_i \\ \vdots \\ y_i, x_i, a_i \\ 
\vdots \end{array} \right] \cdot 
\left[\begin{array}{c} z_1 \\ \vdots \\ z_{2^{2n+1}} \end{array}
\right] = 0.
\end{equation}
If $(x^\prime, y^\prime, a^\prime)$ is a solution of (\ref{keyBalanced})
then also $(y^\prime, x^\prime, a^\prime)$ is a solution, since with
each row $(x_i, y_i, a_i)$ we have also the row $(y_i, x_i, a_i)$
appearing.

Now let $g=(x_1, y_1, a_1)$ and $h=(x_2, y_2, a_2)$ be given elements
from $U^\perp$.

\[ \varphi(g \cdot h) = \left\{ 
\begin{array}{r@{, \quad}l}
(x_1, y_1, a_1) \oplus (x_2, y_2, a_2) & \mbox{if} \; a_2 = 0 \\
(y_1, x_1, a_1) \oplus (x_2, y_2, a_2) & \mbox{if} \; a_2 = 1.
\end{array}
\right.
\]
Since $g\in U^\perp$ also $(y_1, x_1, a_1)\in U^\perp$, so
for all $u\in U$ the following holds:
\begin{eqnarray*}
 \langle \varphi(g\cdot h), u \rangle & =& 
\left\{ \begin{array}{r@{, \quad}l}
\langle (x_1, y_1, a_1), u \rangle + \langle (x_2, y_2, a_2), u \rangle
& \mbox{if} \; a_2 = 0 \\
\langle (y_1, x_1, a_1), u \rangle + \langle (x_2, y_2, a_2), u \rangle
& \mbox{if} \; a_2 = 1 
\end{array}
\right.\\
& =& 0
\end{eqnarray*}
Therefore $g \cdot h \in U^\perp$. Closedness under taking inverses
follows from the fact that elements $u \in W_n$ are either involutions or
$u^3 = u^{-1}$.

"$\Leftarrow$": It is sufficient to show that $U^\perp$ is balanced
since $(U^\perp)^\perp = U$. Without loss of generality we can assume
that $U\subseteq N$, since otherwise there exists $(a, b;1) \in U$ from
which we can deduce $u^{(a, b;1)} = u^t$ for all $u\in U$ and we will
be done.

So we have to show that there exists $(a, b; 1) \in U^\perp$ 
(then $U^\perp$ will by the same argument be balanced and
correspondingly $U$, too). 
Looking at the equations 
\[ 
\left[\begin{array}{c} \vdots \\ x_i, y_i, a_i \\ \vdots 
\end{array} \right] \cdot 
\left[\begin{array}{c} z_1 \\ \vdots \\ z_{2^{2n+1}} \end{array}
\right] = 0
\]
we see that such an element must exist since if $(z_1, \dots,
z_{2^{2n}}, 0)$ is a solution, then $(z_1, \dots,z_{2^{2n}}, 1)$ will
also be a solution.\hfill $\Box$

\noindent
The following corollary summarizes some further properties of the pairing
$\mu$. 

\begin{corollary}
\begin{itemize}
\item[a)] For all subgroups $U\subseteq W_n$
\[ (U^t)^\perp = (U^\perp)^t.\]
\item[b)] Complements of intersections:
\[ (U \cap U^t)^\perp = \langle U^\perp, (U^t)^\perp \rangle
\]
\item[c)] The balanced subgroups of $W_n$ correspond one-to-one to the
balanced subspaces of $\F_2^{2n+1}$.
\item[d)] There is an inclusion-reversing anti-isomorphism $\perp$ on the
lattice of balanced subgroups of $W_n$ which is a Galois
correspondence.  
\end{itemize}
\end{corollary}

\noindent
{\bf Proof:} a) Follows from the fact that $\mu(x, x^\prime) =
\mu(x^t, {x^\prime}^t)$ for all $x, x^\prime \in W_n$. 
b) follows from linear algebra over $\F_2$, c)
is just a reformulation of lemma \ref{balancedness} and d) is obvious.
\hfill $\Box$

%
%

\section{Fourier Transforms for Wreath Products}\label{fourier}

In this section we show how to compute a Fourier transform for the
groups $W_n$ effectively on a quantum computer. We want to do this in
brief since the general recursive method to obtain fast Fourier
transforms on a quantum computer described in \cite{PRB:98} can be
applied directly in case of wreath products $A \wr \z_2$ where $A$ is
an arbitrary abelian $2$-group (for efficient quantum
transforms see also \cite{Hoyer:97}).

\noindent
The recursion of the algorithm follows the chain 
\[ A \wr \z_2 \rhd A \times A \rhd E,
\]
where the second composition factor is the base group. We first want
to determine the irreducible representations of $G:= A\wr {\rm
Z}_2$. Let $G^*$ be the base group of $G$, i.\,e. $G^* = A \times
A$. $G^*$ is a normal subgroup of $G$ of index $2$. Denoting by ${\cal
A} = \{ \chi_1, \ldots, \chi_k\}$ the set of irreducible
representations of $A$ recall that the irreducible representations of
$G^*$ are given by the set $\{\chi_i \otimes \chi_j: i, j = 1,\dots,
k\}$ of pairwise tensor products (see, e.\,g., \cite{Jacobson:89}
section 5.6).

Since $G^* \lhd G$ the group $G$ operates on the representations of
$G^*$ via inner conjugation. Because $G$ is a semidirect
product of $G^*$ with ${\rm Z}_2$ we can write each element $g\in G$
as $g = (a_1, a_2; \tau)$ with $a_1, a_2 \in A$ and we
conclude 
\[ (\chi_1 \otimes \chi_2)^g = (\chi_1^{a_1} \otimes
\chi_2^{a_2})^\tau = (\chi_1 \otimes \chi_2)^\tau,
\]
i.\,e., only the factor group $G/G^* = {\rm Z}_2$ operates via
permutation of the tensor factors. The operation of $\tau$ is to
map $\chi_1 \otimes \chi_2 \mapsto \chi_2 \otimes \chi_1$.

Therefore it is easy to determine the inertia groups (see
\cite{HuppertI:83}, \cite{Beth:84} for definitions) $T_\rho$ of a
representation $\rho$ of $G^*$. We have to consider two cases:
\begin{itemize}
\item[a)] $\rho=\chi_i \otimes \chi_i$. Then $T_\rho = G$ since
permutation of the factors leaves $\rho$ invariant.
\item[b)] $\rho=\chi_i \otimes \chi_j, i\not= j$. Here we have $T_\rho
= G^*$.
\end{itemize}

The irreducible representations of $G^*$ fulfilling a) extend to
representations of $G$ whereas the induction of a representation
fulfilling b) is irreducible. In this case the restriction of the induced
representation to $G^*$ is by Clifford theory equal to the direct sum
$\chi_1 \otimes \chi_2 \oplus \chi_2 \otimes \chi_1$. 

Applying the design principles for Fourier transforms given in
\cite{PRB:98} we obtain the circuits for $\DFT_{W_n}$ in a
straightforward way.  In doing so it is necessary to study the
extension/induction behaviour of representations of $G^*$ since
the recursive formula
\[ \DFT_{G^*} \cdot \bigoplus_{t\in T} \Phi(t) \cdot \DFT_{\z_2}
\]
provides a Fourier transform for $G$. Here $\Phi(t)$ denotes the
extension (as a whole) of the regular representation of $G^*$ to a
representation of $G$ (see \cite{Pueschel:98}, \cite{Beth:84},
\cite{PRB:98}). In case of $W_n$ the transform $\DFT_{G^*}$ is the
Fourier transform for $\z_2^{2n}$ and therefore a tensor product of
$2n$ Hadamard matrices.

\begin{figure}
\input{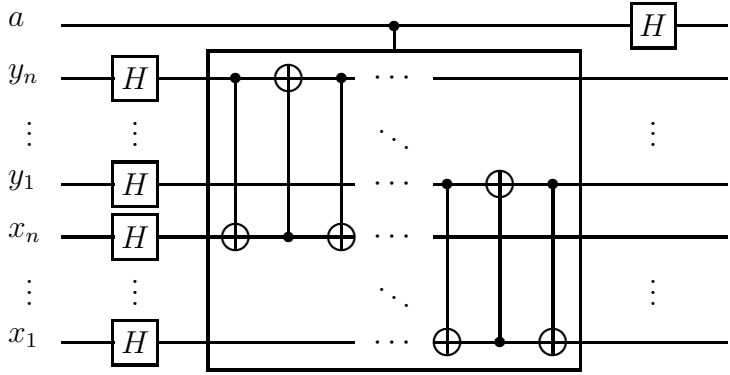}
\vspace*{0.3cm}
\caption{\label{wreathdft}The Fourier transform for $\z_2^n \wr \z_2$}
\end{figure}

The circuits for the case of $W_n$ are shown in figure
\ref{wreathdft}. Quantum circuits in general are built from certain
gate primitives (see \cite{Barenco:95}) and it is clear that the
complexity cost for this circuit is linear in the number of qubits,
since the conditional gate representing the evaluation at the
transversal $\bigoplus_{t\in T} \Phi(t)$ can be realized with $3 n$
Toffoli gates.

Finally we give a slight modification of this circuit by performing
the same matrix $\bigoplus_{t\in T} \Phi(t)$ (which in this case is a
permutation matrix) at the end yielding
\begin{equation}\label{decompsymmetric}
\DFT_{W_n} := \DFT_{G^*} \cdot \bigoplus_{t\in T} \Phi(t) \cdot \DFT_{\z_2}
\cdot \bigoplus_{t\in T} \Phi(t).
\end{equation}

This again decomposes the regular representation of $G$ into
irreducibles and has the advantage to allow a reinterpretation of the
pairing $\mu$ given in section \ref{pairing} for the groups $W_n$:

Multiplying the matrices in (\ref{decompsymmetric}) yields (the
permutation matrix $\Pi$ exchanges the
qubits $x_i$ and $y_i$ for $i=1,\ldots,n$)
\renewcommand{\arraystretch}{1.5}
\[ \DFT_{W_n} = \left( 
\begin{array}{l|l}
H^{\otimes 2 n} & \phantom{-} H^{\otimes 2 n} \cdot \Pi \\
\hline
H^{\otimes 2 n} \cdot \Pi & - H^{\otimes 2 n}
\end{array}\right),
\]
\renewcommand{\arraystretch}{1}
and therefore the matrix entry $\DFT_{[g, h]}$ equals $(-1)^{\mu(g,
h)}$ for all $g, h \in W_n$, where we use the already mentioned
enumeration of the group elements and the pairing $\mu$ defined in
\ref{defpairing}.

%
%

\section{Sampling the Fourier Coefficients}
In this section we address the problem to gain enough information from
the Fourier coefficients under $\DFT_{W_n}$ to find generators for the
$U\cap U^t$ part from factorization (\ref{canonicfactor}).  The
idea is to find generators for the balanced group $\langle U^\perp,
(U^\perp)^t \rangle$ from which we get generators for $U\cap U^t$ by
taking orthogonal complements.

First we want to describe the elements in $U^\perp$:  

\begin{remark}\label{recall}
Let $U\subseteq W_n$. Then one of the following cases holds:
\begin{itemize}
\item[a)] $U^\perp$ consists exclusively of elements of the form $(x,
y; 0)$, i.\,e. $U^\perp= U^\perp \cap N$. Since $U^\perp \cap N$ is a
subgroup of $W_n$ it follows that $U^\perp$ is a group, thus sampling
from an equal distribution over $U^\perp$ will give generators after a
few steps (for an exact analysis see below).
\item[b)] There exists an element $t_0 = (x, y; 1)$ in
$U^\perp$. We can conclude that exactly half of the elements are in
$U\cap N$ and the other half is of the form $t_0 \cdot U\cap N$
(see lemma \ref{halves}).
\end{itemize}
\end{remark}

The following theorem shows that the Fourier transform for the groups
$W_n$ have properties very similar to the abelian case:

\begin{theorem}
Let $\DFT_{W_n}=\sum_{x, y \in W_n} \mu(x, y) \ket{y}\bra{x}$ be the Fourier
matrix. Then for each subgroup $U\subseteq W_n$ we have:

\[ \DFT_{W_n} \frac{1}{|U|} \sum_{x \in U} \ket{x} = 
\frac{1}{|U^\perp|} \sum_{y \in U^\perp} \ket{y}.
\]
\end{theorem}
\noindent
{\bf Proof:} Since 
\begin{eqnarray*}
 \DFT_{W_n} \frac{1}{|U|} \sum_{x \in U} \ket{x} 
& = & \left(\sum_{x, y\in W_n} \mu(x, y) \ket{y}\bra{x} \right)
      \sum_{x\in U} \ket{x}\\
& = & \sum_{y \in W_n}  \sum_{x \in U} \mu(x, y) \ket{y},
\end{eqnarray*}
it suffices to show $\sum_{x\in U} \mu(x, y) = 0\;$ for $\;y\notin
U^\perp$, but this statement is lemma \ref{balancedness}. The other
case $\sum_{x\in U} \mu(x, y) = |U|$ for $y \in U^\perp$ is obvious.

\hfill $\Box$
 
Next we show that sampling yields also information about $U$ in case
we have drawn a coset $g_0 U$ instead of $U$. In case $g_0\in N$ we
indeed sample from $U$, since $N$ acts diagonally in the Fourier basis
with phase factors $\pm 1$, i.\,e.
\[ \DFT_{W_n} \frac{1}{|U|}\sum_{x \in g_0 U} \ket{x} 
= \frac{1}{|U^\perp|}\sum_{y\in U^\perp} \varphi_{g_0, y} \ket{y}
\]
with certain phase factors $\varphi_{g_0, y}$ which depend on $g_0$
and $y$ but are always from $\{\pm 1\}$. Since making measurements
involves taking the squares of the amplitudes we get an equal
distribution over $U^\perp$.

The other case $g_0 \in W_n\setminus N$ leads to an equal distribution
over $(U^t)^\perp$, since an element $g_0 = n t, n \in N$ operates up
to phase factors like $t$ in the Fourier basis and $t$ swaps $(x, y;
a)$ and $(y, x; a)$ when considered as basis vectors of the Fourier basis.

\subsection{Analysis of Sampling}

By the preceding observations we are able to take samples equally distributed
from the sets $U^\perp$ and $(U^t)^\perp$ according to whether $g_0
\in N$ or $g_0 \in W_n \setminus N$. Both cases occur with probability
$1/2$ since $[W_n : N] = 2$. We now have to show that after a few
samples we have found generators for the group $\langle U^\perp,
(U^t)^\perp \rangle$ generated by $U^\perp$ and $(U^t)^\perp$.

We denote the set of sampled elements after the $i$-th measurement by
${\cal E}_i$ and have to give a bound on the probability that ${\cal
E}_i$ generates this group.

\begin{lemma}\label{probab}
For the probability $P$ of finding a set of generators after $i$
samples we have the following estimation
\[ P(\langle {\cal E}_i \rangle = \langle U^\perp, (U^t)^\perp \rangle) 
\geq 1 - 2^{-i/4}.
\]
\end{lemma}
\noindent
{\bf Proof:} We already know that $U^\perp \cap N$ and $(U^t)^\perp
\cap N$ are groups. Also we know from remark \ref{recall} that if
there is one element in $U^\perp$ which is not in $N$ then exactly
half of the elements in $U^\perp$ must be in $N$ and the other half in
$W_n \setminus N$. The same argument holds for $(U^t)^\perp$. Thus in
the worst case we are facing the situation, that with each sample we
fall into one of the boxes
\[
\begin{array}{cccc}
\mbox{\fbox{\vphantom{$(U^\perp \cap N) \cdot t_0$} $U^\perp \cap N$}} &
\mbox{\fbox{$(U^\perp \cap N) \cdot t_0$}} &
\mbox{\fbox{$(U^t)^\perp \cap N$}} &
\mbox{\fbox{$((U^t)^\perp \cap N)\cdot t_0^\prime$}}
\end{array}
\]
(with certain
elements $t_0$ and $t_0^\prime$ from $W_n \setminus N$). The
probability not to have generated $U^\perp \cap N$ and $(U^t)^\perp \cap N$
after $i$ steps is smaller than $2^{-i/4}$. From the other two sets
only one element is necessary to discriminate between the index $1$
and index $2$ case and so the statement follows. \hfill $\Box$

One remark is in order, since it is necessary to have a criterion when
to stop sampling: Arguing like in \cite{BH:97}, suppose the
group generated by ${\cal E}_i$ is to small, i.\,e., after taking
duals we are dealing with $U^\prime \supset U$. Then one of the
generators found must necessarily evaluate to a different value than
the neutral element of $U$ does. This is due to the promise about $f$
and can be checked in polynomial time by comparing the values on all
generators found.

%
%

\section{The Quantum Algorithm}\label{quantumalgo}

Using the results of the preceding sections we can now formulate a
quantum algorithm which solves the hidden subgroup problem for the
non-abelian groups $W_n$. It uses $O(n)$ evaluations of the black box
quantum circuit $f$ and the classical post-computation, which is
essentially linear algebra over $\F_2$, also takes a number of
operations which is polynomial in $n$.

\begin{algorithm}
\begin{enumerate}
\item Prepare the ground state 
\[ \ket{\varphi_{1}} = \ket{0\ldots 0} \otimes \ket{0\ldots 0}
\]
in both registers.
\item Achieve equal amplitude distribution in the first register, for
instance by an application of a Hadamard transform to each qubit:
\[ \ket{\varphi_2} = \sum_{x\in W_n} \ket{x} \otimes
\ket{0\ldots 0}.
\]
(Normalization factors omitted.)
\item Calculate $f$ in superposition and obtain
\[ \ket{\varphi_3} = \sum_{x\in W_n} \ket{x}\ket{f(x)}
\]
\item Measure the second register and obtain 
a certain value $z$ in the image of $f$. In the first register we have
a whole coset $g_0 U$ of the hidden subgroup $U$:
\[ \ket{\varphi_4} = \sum_{f(x)=z} \ket{x}\ket{z} = \sum_{x \in g_0 U}
\ket{x}\ket{z}.
\]
(Like in the case of Simon's algorithm, this step can be omitted.) 
\item Now solve the hidden subgroup problem for the normal
subgroup $N$, which is the base group of $W_n$. This can be done by
application of the standard algorithm for $\z_2^{2n}$ on the first
$2n$ qubits.
\item Application of the Fourier transform on the first register using
the circuit given in section \ref{fourier} transforms the coset into a
superposition of the form $\sum_{x\in U^\perp} \varphi_{g_0, y}
\ket{y}$ in case $g_0\in N$ (with certain phase factors $\varphi_{g_0,
y}$ which depend on $g_0$ and $y$ and are from $\{ \pm 1 \}$).  If
$g_0 \in W_n\setminus N$ we get a superposition over the conjugated
group $\sum_{x\in (U^t)^\perp} \varphi_{g_0, y} \ket{y}$,
\item Now measure the first register. With probability $1/2$ we draw
$g_0$ from $N$ resp. $W_n\setminus N$, i.\,e., we get a superposition
over $U^\perp$ resp. $(U^t)^\perp$ which leads (by performing
measurements) to either equal distribution over $U^\perp$ or equal
distribution over $(U^t)^\perp$.
\item Iterating steps $1.$--$7.$ we generate with high probability (see
lemma \ref{probab}) the group $U^\perp \cap N$ and the group
$(U^t)^\perp \cap N$. 

What is missing are the sets $(U^\perp \cap
N)\cdot t_0$ and $((U^t)^\perp \cap N)\cdot t_0^\prime$ with certain
elements $t_0$ and $t_0^\prime$ not in $N$. It is clear that it is
sufficient to find only one element in one of these two sets, since
then the whole group $\langle U^\perp, (U^t)^\perp\rangle$ will be
generated. But if any, there are many elements of this form in
$U^\perp$ resp. $(U^t)^\perp$ since either there are none of them or
exactly half of the elements of $U^\perp$ resp. $(U^t)^\perp$ is not
in $U^\perp \cap N$ resp. $(U^t)^\perp \cap N$. Summarizing: 

After performing this experiment an expected number of $4 n$ times we
generate with probability greater than $1-2^{-n}$ the group $\langle
U^\perp, (U^t)^\perp\rangle$.
\item By solving linear equations over $\F_2$ it is easy to find
generators for
\[(\langle U^\perp,(U^t)^\perp\rangle)^\perp = U \cap U^t.
\]
After all we get generators for $U = (U \cap N) \cdot (U \cap U^t)$.
\end{enumerate}
\end{algorithm}

%
%

\section{Conclusion and Outlook}

We have presented a family of non-abelian groups for which the hidden
subgroup problem can be efficiently solved on a quantum computer. The
quantum algorithm is followed by a classical post-processing involving
standard linear algebra over the finite field $\F_2$ which can be done
efficiently on a classical machine.

The groups discussed are certain wreath products $W_n$ and our
approach uses a special property of the subgroups of $W_n$ to split
the task of finding generators in two steps: First an abelian hidden
subgroup problem is solved and next the non-abelian Fourier transform
for $W_n$ is used to sample from two sets which in turn allow
reconstruction of the hidden subgroup.

It seems possible to generalize this result to
arbitrary split extensions of the form $\z_2^n \rtimes_\varphi \z_2$
and to examine an approach using representation theory
instead of the pairing used in the paper.

%
%
 
\bibliography{paper}

\begin{thebibliography}{10}

\bibitem{Barenco:95}
A.~Barenco, Ch.~H. Bennett, R.~Cleve, D.~P. DiVincenzo, N.~Margolus, P.~Shor,
  T.~Sleator, J.~A. Smolin, and H.~Weinfurter.
\newblock {Elementary gates for quantum computation}.
\newblock {\em Physical Review~A}, 52(5):3457--3467, November 1995.
\newblock LANL e--preprint quant--ph/9503016.

\bibitem{Beth:84}
Th. Beth.
\newblock {\em {Methoden der schnellen Fouriertransformation}}.
\newblock Teubner, 1984.

\bibitem{BH:97}
G.~Brassard and P.~{H\o yer}.
\newblock {An Exact Polynomial--Time Algorithm for Simon's Problem}.
\newblock In {\em {Proceedings of Fifth Israeli Symposium on Theory of
  Computing and Systems}}, pages 12--33. ISTCS, IEEE Computer Society Press,
  1997.
\newblock LANL preprint quant--ph/9704027.

\bibitem{EH:98}
M.~Ettinger and P.~{H\o yer}.
\newblock {On Quantum Algorithms for Noncommutative Hidden Subgroups}.
\newblock LANL e--preprint quant-ph/9807029, 1998.

\bibitem{Hoyer:97}
P.~H{\o}yer.
\newblock {Efficient Quantum Transforms}.
\newblock LANL preprint quant--ph/9702028, February 1997.

\bibitem{HuppertI:83}
B.~Huppert.
\newblock {\em {Endliche Gruppen}}, volume~I.
\newblock Springer, 1983.

\bibitem{Jacobson:89}
N.~Jacobson.
\newblock {\em {Basic Algebra II}}.
\newblock Freeman and Company, 1989.

\bibitem{JK:82}
G.~James and A.~Kerber.
\newblock {\em {The Representation Theory of the Symmetric Group}}.
\newblock Cambridge University Press, 1982.

\bibitem{Jozsa:98}
R.~Jozsa.
\newblock {Quantum Algorithms and the Fourier Transform}.
\newblock {\em Proc. R. Soc. Lond. A}, 454:323--337, 1998.

\bibitem{ME:98}
M.~Mosca and A.~Ekert.
\newblock {The Hidden Subgroup Problem and Eigenvalue Estimation on a Quantum
  Computer}.
\newblock In {\em Proceedings 1st NASA International Conference on Quantum
  Computing \& Quantum Communications}, LNCS 1509. Springer, 1998.

\bibitem{Pueschel:98}
M.~P{\"u}schel.
\newblock {\em {Konstruktive Darstellungstheorie und Algorithmengenerierung}}.
\newblock PhD thesis, Univ. Karlsruhe, Informatik, 1998.

\bibitem{PRB:98}
M.~P{\"u}schel, M.~R{\"o}tteler, and Th. Beth.
\newblock {Fast Quantum Fourier Transforms for a Class of non-abelian Groups}.
\newblock LANL e--preprint quant-ph/9807064.

\bibitem{Shor:94}
P.~W. Shor.
\newblock {Algorithms for Quantum Computation: Discrete Logarithm and
  Factoring}.
\newblock In {\em Proceedings of the 35th Annual Symposium on Foundations of
  Computer Science}, pages 124--134. Institute of Electrical and Electronic
  Engineers Computer Society Press, November 1994.

\bibitem{Simon:94}
D.~R. Simon.
\newblock On the power of quantum computation.
\newblock In {\em Proceedings of the 35th Annual Symposium on Foundations of
  Computer Science}, pages 116--123, Los Alamitos, CA, 1994. Institute of
  Electrical and Electronic Engineers Computer Society Press.

\end{thebibliography}
 
\bibliographystyle{plain}  

\end{document}